# Deep recommender engine based on efficient product embeddings neural pipeline

Laurentiu Piciu
*Lummetry.AI*
Bucharest, Romania
laurentiu@lummetry.ai

Andrei Damian
*Lummetry.AI*
Bucharest, Romania
andrei@lummetry.ai

Nicolae Tapus
*University Politehnica of Bucharest*
Bucharest, Romania
ntapus@cs.pub.ro

Andrei Simion-Constantinescu
*Knowledge Investment Group*
Bucharest, Romania
andrei.simion@kig.com

Bogdan Dumitrescu
*High-Tech Systems & Software*
Bucharest, Romania
bogdan.dumitrescu@htss.ro

*Abstract* — Predictive analytics systems are currently one of the most important areas of research and development within the Artificial Intelligence domain and particularly in Machine Learning. One of the "holy grails" of predictive analytics is the research and development of the "perfect" recommendation system. In our paper, we propose an advanced pipeline model for the multi-task objective of determining product complementarity, similarity and sales prediction using deep neural models applied to big-data sequential transaction systems. Our highly parallelized hybrid model pipeline consists of both unsupervised and supervised models, used for the objectives of generating semantic product embeddings and predicting sales, respectively. Our experimentation and benchmarking processes have been done using pharma industry retail real-life transactional Big-Data streams.

## I. Introduction

Recommender systems are by far one of the most important areas where machine learning in conjunction with Big Data are applied with proven success. The goal is to find what is likely to be of interest to a certain customer or group of customers and to provide personalized services to them. The focus in this area is quite intensive due to the need of finding the "*perfect*" recommendation system that will allow retailers to structure their offer including sales strategy and marketing campaigns very well in order to optimize consumers experience.

This paper proposes a state-of-the-art deep neural model for the multi-task objective of determining product complementarity & similarity, which leads to diversified market baskets and finally to sales prediction. The pharmaceutical industry has been proposed as an experimental environment in our research. As a result, we used the advantage of a Big-Data store that provided sequential transaction information. The data has been provided by a successful pharma retailer and a little background is necessary for the application of the proposed approach in this particular vertical.

*Recommender systems in the pharma industry.* An important aspect related to the challenge of constructing pharma retail recommender systems resides in the transactional flow of this particular industry. In pharma retail, we have both traceable (using loyalty cards) customers and unknown customers. Nevertheless, the recommender system should be able to predict various targets such as a best potential offer for a market basket starting from a product (or products), automatic inference of consumer health/medical condition needs, product complementarity, similarity, and actual product cannibalization. All of these mentioned use-cases within the pharma industry have to be tackled either if we have time-series data available for a particular customer (time-series based profiling) or time-series and customer-agnostic predictive analytics. To be more specific, in most other retail verticals, the focus is on customer loyalty based on historical data and collaborative filtering [1] with little focus on history-agnostic approaches. Even more, most retail verticals do not have meta-products or relevant meta-information similar to these examples from pharma retail vertical: target potential disease condition, target physiological need, target body part, etc. These presented use-cases are included in our current and near-future focus in the area of customer-oriented and customer-agnostic recommender systems.

Our work presents a way to extend capabilities of recommender systems by learning low-dimensional vector space representation of products and users – embeddings – used in the following stage for sales prediction and introduce this process as an actual step in the overall end-to-end deep learning pipeline. The final objective, as previously mentioned, is to address several use-cases starting with product complementarity and similarity baskets.

The taxonomy of the proposed pipeline architectural view can be divided into a use-case approach and a model-based design overview. From the perspective of model-oriented architecture our pipeline can be divided into two separate general models with end-to-end learning capabilities: the early stage of product semantic analysis and later stage of product sales regression. The key point within the initial stages of our pipeline system is learning product and customer feature vector embeddings. Our work is analogous to Word2Vec [2] and GloVe [3] approaches which are used to learn linguistic regularities and semantic information for natural language processing. The general approach within the initial stage is based on analyzing each sequence from the transactions database, choosing $c$ products around a target product $p_i$ within the same transactional *document* in the same way an NLP system would analyze text semantics within same *sentences*. The end result will reveal that the embeddings generate *clusters* of similar products within a latent space of products. As a result, through this approach, we can identify a set of *k* items that will be of interest to a certain customer. Moreover, besides the product similarity, the resulting embeddings will bring to light other important insights such as if a product is not available anymore, it can be replaced with other two products whose *combined embeddings* will be very close to the initial product feature vector. Our product embeddings can be directly used in order to determine pharmacy product similarity assessment based on cosine similarity. A more advanced use, also derived from known neural-linguistic models approaches and experimentation, is that of generating ***concept*** products that capture specific needs and answer advanced queries such as "What natural remedies are good for back pain as it is Vitamin C for simple cold?" that actually translates in a more domain-specific query such as "What products are complementary to product A as it is product Z for product X?".

The main need for business predictive analytics models that can accurately recommend market baskets generating in this process products and users' *meaningful* latent semantic spaces comes from the increasing evolution of online advertising approaches. Targeting accurately the users in order to meet their real actual demands is a critical aspect that drives consumers to visit again the sites or to use the applications without being "bombarded" by uninteresting recommendations. ***Consequently, the business success of retailers is now strongly related to their user retention capability, which is a result of their ability to generated consumer-tailoring content for every user-experience moment.***

Our work includes also a deep neural model which uses the consumer and product feature vectors and directly attempts to predict sales information. As we mentioned, this is a crucial aspect for all companies if they want to set up an efficient marketing campaign.

To the best of our knowledge, this work represents the first study that offers an end-to-end recommendation solution in the pharmaceutical field.

An important observation regarding our industrial research and experimental development is that our focus has been exclusively on behavioral analytics with marginal direct impact on sales forecasting that is proposed for the next steps of our research. As a result, the presented directed acyclical graphs follow simple baseline design patterns.

In terms of paper structure, Section II will present the approaches that our work relates while in Section III we present our proposed P2E (Product-to-Embeddings), U2E (User-to-Embeddings) and ProVe (Product Vectors) models, together with our proposed baseline deep neural model for final stages of sales prediction. In Section IV are presented the results of the proposed recommender system using a large pharmaceutical transactional database. Finally, Section V will emphasize the conclusions of this work and the planned future development directions.

## II. RELATED WORK

Our work relates with several approaches either derived from Natural Language Processing or from classic methods that address the problem of recommendation systems.

### A. Traditional Approaches

Existing methods for recommender systems can easily be categorized into **collaborative filtering** (CF) methods [1], [4], [5] and content-based methods [6], which make use of the user or product content profiles. Collaborative filtering is based on user-item interactions and predicts which products a user will most likely be interested in by exploiting purchase behavior of users with similar interests or by using user's interaction with other products. CF methods popularity is based on the fact that they can discover interesting associations between products and do not require the heavy knowledge collection needed by content-based methods. To mitigate the cold-start problem, which CF methods suffer from, matrix factorization-based models [7] have been developed and now they are very popular after their success in the Netflix competition.

**Matrix factorization** [7], [8] for collaborative filtering models can approximate a sparse user-item interaction matrix by learning a latent representation of users and items using SVD or stochastic gradient descent which give the optimal factorization that globally minimizes the mean squared prediction error over all user-item pairs.

### B. Neural NLP Models

In several Natural Language Processing (NLP) tasks (e.g., computing similarity between two documents), learning linguistic regularities, as well as semantic information is essential. Therefore, a mathematical model has been developed by Mikolov et al. [2] (**Word2Vec**), which can be used for learning high-quality low-dimensional word embeddings from huge datasets and huge vocabularies, using two architectures of neural networks: continuous bag-of-words (CBOW) and skip-gram (SG).

This powerful and efficient model takes advantage of the word order in the text documents, explicitly modeling the assumption that closer words in a context window are statistically more dependent. In the SG architecture, the objective is to predict a word's context given the word itself, whereas the objective in the CBOW architecture is to predict a word given its context.

**GloVe** [3] is another successful model for learning multi-dimensional vector representations of words by combining the advantages of two major model families in the NLP literature: global matrix factorization and local context window methods (Word2Vec). They consider the primary source of information available about a corpus of words being the word-word co-occurrence counts which are used to train the fine-grained word embeddings. Explicitly, the ratio of the co-occurrence probabilities of two words (rather than their co-occurrence probabilities themselves) is what contains the information encoded as word embeddings.

## III. Approach

### A. From NLP to Recommender Systems

Traditionally, in NLP applications, each word is represented as a feature vector using a one-hot encoding where a word vector has the same length as the size of the vocabulary. Our first approach was to create a corpus of words from all pharmaceutical prospectuses and to encode each product, using **tokenized-word features,** such as feature $i$ is 1 if the word $i$ appears in the prospectus of the given product and 0 otherwise. Then, we created a model based on XGBosst Regression Trees [9] which predicted sales for segments of users with a mean 74% R2 score. Nevertheless, this approach suffers from several drawbacks such as high dimensionality and sparsity of the input and low-viability for the implementation of our second target of predicting market baskets.

### B. Products Embeddings

Considering the big improvement word embeddings brought to NLP domain, **we were confident that from language models to product business analytics is a very slight difference**.

To address the task of finding complementarity/similarity between products and diversified market baskets for a certain customer, we proposed to learn representations of products in low-dimensional space, using the Big-Data sequential transaction system provided by the pharma retailer, which was used also for computing the word-word co-occurrence counts.

More specifically, we developed two models (**P2E** and **ProVe**) inspired by the NLP models Word2Vec and GloVe**.** Our first model uses sequential transactions. The transaction system can be represented as a set $S$, where:

- $S = (t_1, t_2, t_3, ..., t_M), M = $ total number of transactions/market baskets/receipts
- $t_i = (p_{i_1}, p_{i_2}, ..., p_{i_{T_i}}), T_i = $ total number of products on $i$-th receipt.

Our objective is to find $D$-dimensional real-valued representations $w_{p_i} \in \mathbb{R}^D$ of each product $p_i$ such that they lie in a latent vector space. The general approach within this initial stage is based on analyzing each transaction, predicting each target product $p_i$ using all context products (all products that are on the same receipt) (Figure 1). **If there is only one product in a market basket, it is skipped since the transaction does not contain complementarity information**.

Therefore, the objective function of CBOW architecture is defined as follows

$$J = \frac{1}{M} \sum_{i=1}^{M} \log \mathbb{P}(p_i \mid p_{i-c}, ..., p_{i-1}, p_{i+1}, ..., p_{i+c}) \quad (1)$$

where probability $\mathbb{P}(p_i \mid p_{i-c}, ..., p_{i-1}, p_{i+1}, ..., p_{i+c})$ of predicting the current product $p_i$, given a product context is defined using the *softmax* function (2):

$$\mathbb{P}(p_i \mid p_j) = \frac{e^{w_{p_j}^T w_{p_i}}}{\sum_{k=1}^{P} e^{w_{p_j}^T w_{p_k}}} \quad (2)$$

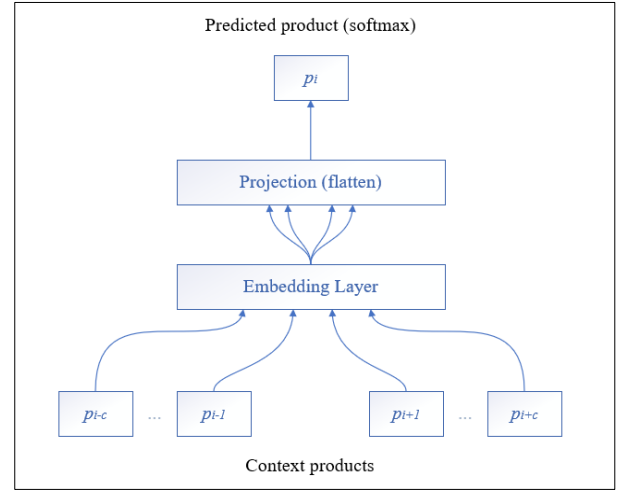

Figure 1 – P2E CBOW architecture

The original Word2Vec model proposes a sampled-softmax loss [2] which is a computationally efficient approximation of the full softmax. Given the fact that our models are trained in a high-performance computing environment, we used the non-approximated version which resulted in better results. Simultaneously, the projection layer computes a concatenation of all $c * D$ embeddings, instead of summation (Figure 1). This aspect leads to more computation for the softmax layer, but also to better accuracy.

The second model (**ProVe**) seeks to learn low-dimensional product embeddings using the ratio of co-occurrence probabilities of two products. Therefore, the model generates two sets of product vectors $W$ and $\widetilde{W}$ and it should minimize the weighted least squares objective $J$ defined as follows:

$$J = \sum_{i,j=1}^{P} f(X_{ij}) \, (w_{p_i}^T \widetilde{w}_{p_j} + b_i + \tilde{b}_j - \log X_{ij})^2 \quad (3)$$

The product-product co-occurrence score measures how often two products are bought together in the same market basket (4).

$$X_{ij} = X_{ij} + \frac{1}{d(p_i, p_j)} \quad (4)$$

where $p_j$ represents a product that is in the context of the product $p_i$ for each receipt and $d$ represents the distance between these products in a context window (receipt).

For both **P2E** and **ProVe**, we applied the K-means algorithm [10] on the resulting embeddings. This is our first simple approach for the objective of determining the *concept vectors* which should capture all the *physiological needs* of a certain product. Again, in the particular case of pharmaceutical products, the concept vectors may capture, for example, information about the position of the *'flu'* concept in the semantic latent space, which leads to obtaining full similarity and/or complementarity in the generation of the market basket.

The market baskets are efficiently created starting from a certain product by using the **cosine similarity** (5) between it and all product vectors. **Algorithm 1** defines the methodology used for market baskets definition.

$$similarity(p_i, p_j) = \frac{w_{p_i} \cdot w_{p_j}}{||w_{p_i}||_2 \, ||w_{p_j}||_2} \quad (5)$$

---

**Algorithm** 1 *MarketBasket*(*Product, EmbeddingSpace, k*)

    e $\Leftarrow$ get embedding of Product from *EmbeddingSpace*
    cos_dist $\Leftarrow$ *compute_cos_sim(e, EmbeddingSpace)*
    top_k $\Leftarrow$ choose top *k* closest products based on cos_dist
    Eliminate products that have the same *concept vector* as *Product*
    market_basket $\Leftarrow$ set of all complementary products
**Return** market_basket

---

Our approach is similar to current state-of-the-art recommender systems (Grbovic et al. [11], Vasile et al. [12]) that use a multi-dimensional representation of an ecosystem's entities (products, services, etc.).

*C. Users Embeddings*

Motivated by the doc2vec algorithm proposed by Le et al. [13], which jointly optimizes both word embeddings and the global context of the entire document, we also employed this methodology in order to create a latent multi-dimensional semantic space for both the users and the products representations.

Considering that a receipt (market basket) is defined by the products that are bought and by the "global context" (user/customer), the P2E architecture is modified (Figure 2) for the objective of jointly learning users and products embeddings. Therefore, the cost function that should be minimized is defined as follows:

$$J = \frac{1}{M} \sum_{i=1}^{M} \log \mathbb{P}(p_i \mid u_j, p_{i-c}, \dots, p_{i-1}, p_{i+1}, \dots, p_{i+c}) \quad (6)$$

where probability $\mathbb{P}(p_i \mid u_j, p_{i-c}, \dots, p_{i-1}, p_{i+1}, \dots, p_{i+c})$ is defined using the *softmax* function (*2*).

The customer embedding space is created **exclusively using transactional information** (no personal information). Therefore, this latent space encodes the similarity between customers according to their buying behavior. The latent vectors corresponding to each person are already used in pharmaceutical commercial applications for various objectives:

- Segmentation of customers according to their tastes (which products they have bought);
- Individual marketing campaigns;
- Products propensity-to-buy prediction.

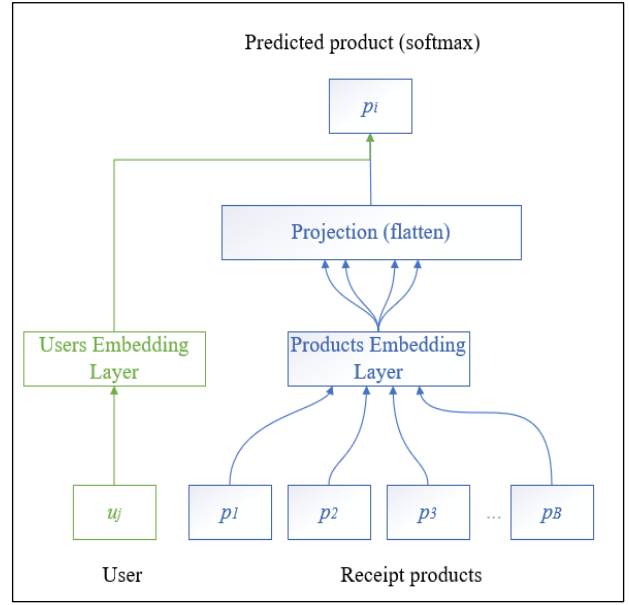

Figure 2 – U2E CBOW architecture

The training data used for the user embeddings model jointly trained with the product embeddings has over 4 million user data ranging from customers with **a transaction** up to customers with **2000 transactions.** The actual user embeddings calculation can be evaluated based on the number of transactions per each user (the more the better). On average, our model optimizes the user embeddings (starting from an initial random uniform allocation) on more than 60% of the population.

In order to evaluate customer (user) embeddings, we must first select the ones that have passed the minimal number of transactions in order to ensure that the backpropagation-based optimization process has managed to sufficiently nudge the initial randomly uniform embedding position and thus assign a viable location to the customer in the latent space. Following this first step, we can either use the user embeddings in order to train shallow or deep models for customer-product propensity to buy inference or use simple analytics approaches in order to compare customer average product baskets for customers that reside in the same area of the embedding space.

*D. Sales Regression Model*

The last model in the proposed deep neural pipeline predicts customer buying propensities for any product in the ecosystem based on their optimized embedding spaces, in which every entity has a well-determined position "extracted" from the sequential transactional information.

This regression model supports any kind of sales aggregation (seasonal, yearly, etc.). As a result, it will generate the customer-product propensity-to-buy for the next period. From a business point of view, the model is involved in pharmaceutical marketing campaigns which offer discounts for customers based on their propensities-to-buy for a certain period.

For this particular model, we researched and developed a fully-connected deep neural network [14] with 3 layers. This neural network has two inputs: one for user embeddings and one for product embeddings and the readout layer has a single neuron which is a regressor (7) and predicts totally supervised how much will spend the customer for a certain product.

$$J = \frac{1}{M}\sum_{i=1}^{M}(f(u_i, p_i) - y_i)^2 \quad (7)$$

*E. End-to-end Pipeline Formalization*

Our approach manages to build embeddings for products and users from transactional pharmaceutical data. The resulting embeddings are used to predict the propensity-to-buy of each customer to any product. The obtained prediction is used for recommendation, together with information extracted the market basket prediction. An end-to-end pipeline formalization is presented below:

- $\{w_{p_1}, w_{p_2}, \ldots, w_{p_n}\} = f(p_1, p_2, \ldots, p_n)$, where $n$ is the total number of products and $f$ is the product embedding function optimized using (*1*) via backpropagation, which returns the $D_1$-dimensional real-valued representations $w_{p_i} \in \mathbb{R}^{D_1}$;

- $\{w_{u_1}, w_{u_2}, \ldots, w_{u_m}\} = g(u_1, u_2, \ldots, u_m)$, where $m$ is the total number of users and $g$ is the user embedding function optimized using (*6*) via backpropagation, which returns the $D_2$-dimensional real-valued representations $w_{u_j} \in \mathbb{R}^{D_2}$;

- $\{p_{compl_1}, p_{compl_2}, \ldots, p_{compl_k}\} = MB(p_i, f, k)$, where $p_i$ is the product for which is needed to find complementary products, $f$ is the optimized product embedding function, $k$ is the dimension of the market basket and $MB$ is the function that computes the set of all complementary products based on similar embeddings;

- $total\_amount = h(u_j, p_i)$, where $h$ is the function optimized using (*7*) via backpropagation which returns the propensity-to-buy of the customer $u_j$ to the product $p_i$.

## IV. EXPERIMENTS AND RESULTS

The pipeline was trained, validated and tested using an existing *Big-Data* sequential transaction system comprising more than 200 million purchases made by 4.3 million users, involving about 27,000 unique pharmaceutic products within a 24 months' time-frame. All the models were trained on a high-performance computing (HPC) environment using CUDA [15] kernels that are deployed on an NVIDIA Pascal 5000 based GPU card.

*A. Market Basket Experiment*

For the first objective of creating market baskets, we trained our product and user embeddings models – P2E, ProVe and U2E – during 50 epochs and using Adagrad [16] optimizer with learning rate = 1 and initial accumulator value = 0.1. Each epoch lasted 1 hour and 35 minutes on the mentioned HPC.

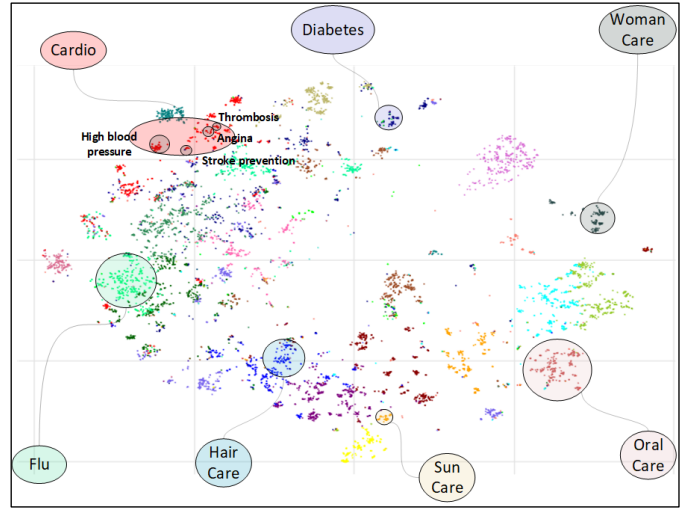

Figure 3 – Pharmaceutical products "meta-map"

In order to visually analyze the product embeddings, we used the classic approach and transformed them in a two-dimensional "meta-map" (Figure 3) employing t-Distributed Stochastic Neighbor Embedding (t-SNE) technique [17]. In *Figure 3* we have highlighted several different regions that actually contain **strong semantic meanings**: cardiovascular products (which can be used for high blood pressure, thrombosis, angina or stroke prevention), products used for woman care, products used against flu, etc. The colors represent the product clusters discovered using the K-means algorithm.

*B. Sales Regression Experiments*

For the sales regression objective, we used a straight-forward simple fully-connected deep neural network (160-80-1) presented as a graph in Table 1. We structured a 1-year transactional database such that each observation was defined by a pair customer-product and the total amount of money spent by the customer on that product during a year.

Formally, we can define the dataset as a pair $([X_1, X_2], y)$, where $X_1, X_2$ represents the user and product IDs and $y$ represents the target – the sales generated by the customer on that product.

The fully-connected deep neural network (160-80-1) was trained during 35 epochs using Adam [18] optimizer with learning rate = 0.0025 and batch size = 512. Each epoch lasts almost 90 seconds on the mentioned HPC.

TABLE 1 FC DEEP NEURAL NETWORK FOR SALES PREDICTION

| Layer (type) | Output Shape | Connected to |
|---|---|---|
| input_user (*InputLayer*) | (None, 1) | |

| Layer (type) | Output Shape | Connected to |
|---|---|---|
| input_prod (*InputLayer*) | (None, 1) | |
| emb_user (*Embedding*) | (None, 32) | input_user |
| emb_prod (*Embedding*) | (None, 128) | input_prod |
| dense_input (*Concatenate*) | (None, 160) | emb_user<br>emb_prod |
| fc_layer_1 (*Dense*) | (None, 160) | dense_input |
| fc_layer_2 (*Dense*) | (None, 80) | fc_layer_1 |
| readout_layer (*Dense*) | (None, 1) | fc_layer_2 |

We have also experimented early in our research with hand-engineered features based on a bag-of-words approach on product descriptions. Nevertheless, the product embeddings proved much more efficient and scalable both in terms of resource and performance.

The **baseline** for our sales regression model was our model based on XGBosst Regression Trees and hand-engineered features (III.A) which obtained a **74% r2-score**. This model cannot be used in production-grade systems due to very high data sparsity. Another main drawback of this model is its lack of knowledge about the powerful embedding spaces created using our P2E and U2E models.

We defined four different experiments (Table 2) whose purpose was to determine the effect of continuing/not continuing the optimization of the embeddings based on the sales registered for each customer. **The optimization of the latent spaces in the sales prediction process also encodes other meta information such as:**

- How much each customer spends;
- Who are the customers that have the propensity to buy more;
- Price of the products.

Consequently, the deep neural model is a function $f(u_i, p_i)$ that plays two main roles: predicts the total sale amounts given two entities (customer/user, product) encoded in their latent spaces and optimizes the embeddings according to the sales registered for each customer/user.

TABLE 2 EMBEDDINGS OPTIMIZATION EXPERIMENTS

| Experiment ID | Continue opt. user embeddings | Continue opt. prod embeddings | R2 score |
|---|---|---|---|
| 1 | ✗ | ✗ | 84% |
| 2 | ✗ | ✓ | 76% |
| 3 | ✓ | ✗ | **91%** |
| 4 | ✓ | ✓ | 88% |

## V. CONCLUSIONS AND FURTHER WORK

### A. Conclusions

The presented work describes the research and development steps for an end-to-end recommender system which is part of a commercial application that is already being used by a pharmaceutical retail company in their sales strategy and marketing campaigns. Due to the following main approaches:

- a workflow that generates completely unsupervised products' and users' semantical information (embeddings) using only sequential transactional data;
- a technique that uses the products' embeddings to compute the concept vectors of the ecosystem (in our particular case, the pharmaceutical ecosystem) in order to generate complementary market baskets;
- an approach that uses the already optimized embeddings and generates year-to-year sales prediction, while, simultaneously, enables further optimization of the embeddings in order to capture the *amount*-related information along with the transactional patterns;

Finally, we can conclude that our work manages to innovate the area of predictive analytics with an emphasis on consumer behavior analytics.

### B. Underway and Future Improvements

It remains to explore what are the benefits of the current state-of-the-art neural networks in sequence processing (recurrent neural networks – particularly, long short-term memory [19]) for recommender systems. More precisely, we are currently researching a multi-stage (hierarchical) recurrent model that is able to process all the transactions of a customer and finally predict which basket is most likely to be bought next by that particular user.

We believe that this proposed future model will prove a powerful tool as it will encode the seasonality and temporality of the consumer actions within the user embedding latent space, alongside the product ones.